\documentclass[journal=nalefd,manuscript=article]{achemso}
\usepackage[version=3]{mhchem} 
\usepackage{graphicx}
\usepackage{capt-of}
\usepackage[english]{babel}
\usepackage[utf8]{inputenc}
\usepackage{siunitx}
\usepackage{sidecap}
\usepackage{subfigure}
\usepackage{amssymb}
\usepackage[dvipsnames]{xcolor}
\usepackage{xspace,colortbl,rotating}
\usepackage{amsfonts}
\usepackage{tabulary}
\usepackage{hyperref}
\hypersetup{
    colorlinks=true,
    linkcolor=Blue,
    citecolor = Blue,
    filecolor=red,      
    urlcolor=Cerulean
    }
\author{Sonka Reimers}
    \affiliation{School of Physics and Astronomy, University of Nottingham, Nottingham NG7 2RD, United Kingdom}
    \email{Sonka.Reimers@nottingham.ac.uk}
    \alsoaffiliation{Diamond Light Source, Chilton OX11 0DE, United Kingdom}
    \author{Dominik Kriegner}
    \affiliation{Institut f\"ur Festk\"orper- und Materialphysik and W\"urzburg-Dresden Cluster of Excellence ct.qmat, Technische Universit\"at Dresden, 01062 Dresden, Germany}
    \alsoaffiliation{Institute of Physics, Czech Academy of Sciences, 162 00 Praha 6, Czech Republic}
    \author{Olena Gomonay}
	\affiliation{Institut f\"ur Physik, Johannes Gutenberg Universit\"at Mainz, 55099 Mainz, Germany}
    \author{Dina Carbone}
    \affiliation{MAX IV Laboratory, Lund University, 22100 Lund, Sweden}
    \author{Filip Krizek}
    \affiliation{Institute of Physics, Czech Academy of Sciences, 162 00 Praha 6, Czech Republic}
    \author{Vit Nov\'{a}k}
    \affiliation{Institute of Physics, Czech Academy of Sciences, 162 00 Praha 6, Czech Republic}
    \author{Richard P. Campion}
    \affiliation{School of Physics and Astronomy, University of Nottingham, Nottingham NG7 2RD, United Kingdom}
   	\author{Francesco Maccherozzi}
    \affiliation{Diamond Light Source, Chilton OX11 0DE, United Kingdom}
    \author{Alexander Bj{\"o}rling}
    \affiliation{MAX IV Laboratory, Lund University, 22100 Lund, Sweden}
	\author{Oliver J. Amin}
    \affiliation{School of Physics and Astronomy, University of Nottingham, Nottingham NG7 2RD, United Kingdom}
    \author{Luke X. Barton}
    \affiliation{School of Physics and Astronomy, University of Nottingham, Nottingham NG7 2RD, United Kingdom}
    \author{Stuart F. Poole}
    \affiliation{School of Physics and Astronomy, University of Nottingham, Nottingham NG7 2RD, United Kingdom}
    \author{Khalid A. Omari}
    \affiliation{School of Physics and Astronomy, University of Nottingham, Nottingham NG7 2RD, United Kingdom}
    \author{Jan Michalička}
    \affiliation{Central European Institute of Technology, Brno University of Technology, 612 00, Brno, Czech Republic}
    \author{Ondřej Man}
    \affiliation{Central European Institute of Technology, Brno University of Technology, 612 00, Brno, Czech Republic}
	\author{Jairo Sinova}
	\affiliation{Institut f\"ur Physik, Johannes Gutenberg Universit\"at Mainz, 55099 Mainz, Germany}
	\author{Tom\'{a}\v{s} Jungwirth}
    \affiliation{Institute of Physics, Czech Academy of Sciences, 162 00 Praha 6, Czech Republic}
    \affiliation{School of Physics and Astronomy, University of Nottingham, Nottingham NG7 2RD, United Kingdom}
    \author{Peter Wadley}
    \affiliation{School of Physics and Astronomy, University of Nottingham, Nottingham NG7 2RD, United Kingdom}
	\author{Sarnjeet S. Dhesi}
    \affiliation{Diamond Light Source, Chilton OX11 0DE, United Kingdom}
    \email{Sarnjeet.Dhesi@diamond.ac.uk}
    \author{Kevin W. Edmonds}
    \affiliation{School of Physics and Astronomy, University of Nottingham, Nottingham NG7 2RD, United Kingdom}
    \email{Kevin.Edmonds@nottingham.ac.uk}

\title[]{{Defect-driven antiferromagnetic domain walls in CuMnAs films}}

\begin{document}
\newpage
\begin{abstract}
Efficient manipulation of antiferromagnetic (AF) domains and domain walls has opened up new avenues of research towards ultrafast, high-density spintronic devices.
AF domain structures are known to be sensitive to magnetoelastic effects, but the microscopic interplay of crystalline defects, strain and magnetic ordering remains largely unknown. 
Here, we reveal, using photoemission electron microscopy combined with scanning X-ray diffraction imaging and micromagnetic simulations, that the AF domain structure in CuMnAs thin films is dominated by nanoscale structural twin defects. We demonstrate that microtwin defects, which develop across the entire thickness of the film and terminate on the surface as characteristic lines, determine the location and orientation of \SI{180}{\degree} and \SI{90}{\degree} domain walls. The results emphasize the crucial role of nanoscale crystalline defects in determining the AF domains and domain walls, and provide a route to optimizing device performance.

\end{abstract}
\maketitle

\section{Introduction}
A key goal of spintronics is the development of
high-speed, high-density data storage devices that are robust against 
magnetic fields. Antiferromagnetic (AF) materials 
offer a route to realising these goals since they exhibit 
intrinsic dynamics in the THz-regime, 
lack magnetic stray fields and can be electrically 
switched\cite{Jungwirth2016, Baltz2018,Zelezny2014,Wadley16}.
Moreover, AF order is exhibited in a wide range of materials 
compatible with the properties of insulators, semiconductors and metals.
Electrical switching has been achieved in several AF systems, with the resulting current-induced domain modifications attributed to spin-orbit torques or thermomagnetoelastic effects \cite{Wadley16, Baldrati2019, kaspar21, meer21}. Spin-orbit torque manipulation
of AF domains was first achieved using orthogonal current
pulses to induce \SI{90}{\degree} rotations of the AF order parameter, but more recently
current-polarity dependent switching of AF order has been achieved
and ascribed to domain wall motion \cite{Wadley2018}.  
AF domains and domain walls are therefore the 
building blocks of AF spintronics, but 
pinning can limit device performance whilst 
creep affects long-term memory stability.  
In ferro- and ferrimagnets, magnetic domain formation has been extensively 
studied for decades and is well-known to be largely governed by the minimization of the demagnetizing field energy \cite{hubert1998, McCord_2015}. 
On the other hand, domain formation in fully compensated 
antiferromagnets remains largely unexplored.

Domain morphologies in AF thin films vary considerably 
with thickness and nanostructure shape which has been ascribed
to strain effects, although evidence for a direct relationship is lacking \cite{Menon2011, Scholl00,Folven2010,Bezencenet11,Folven2015a}.
To date, device concepts have considered an ideal 
AF spin lattice \cite{Zelezny2014,Khymyn2017, Gomonay2018_TerahertzDetector},
but high-resolution AF domain imaging has revealed pronounced 
non-uniformities and pinning effects during domain switching \cite{Grzybowski2017,Wadley2018,Bodnar2018,Baldrati2019,Janda2020,kaspar21}. 

The metallic antiferromagnet CuMnAs is a focus of AF spintronics research due to its favorable crystal symmetry for spin-orbit torque switching. In CuMnAs thin films, elongated microtwins and atomically 
sharp anti-phase boundaries \cite{Krizek2020} have recently been identified
as the most prominent defects. 
The anti-phase boundaries have been associated with atomically sharp \SI{180}{\degree} AF domain walls \cite{krizek2020atomically}. 
The microtwin defects, which are the focus of this work, are shown to have a dramatic influence on the AF domain configuration in CuMnAs thin films. We show 
that microtwin defects largely control the domain structure by generating pinned \SI{90}{\degree} domain walls and confining \SI{180}{\degree} domain walls.

\section{Results}
The \SI{50}{\nano\meter} thick CuMnAs(001) films were grown epitaxially on GaP(001) \cite{Krizek2020}. The CuMnAs layer is a collinear antiferromagnet with a N\'eel temperature of \SI{485}{K} \cite{Hills2015} and a tetragonal crystal structure ($a=b=$\SI{3.853}{\angstrom}, $c=$\SI{6.278}{\angstrom}). Close lattice-matching along the half-diagonal of the cubic GaP substrate unit cell ensures fully strained epitaxial growth  with low mosaicity \cite{Wadley13}. The spin axes align in the $(001)$-plane, \textit{i.e.} in the plane of the film \cite{wadley15} due to a strong magnetocrystalline anisotropy. In the following the crystallographic axes refer to the orientation of the CuMnAs crystal.
 
\begin{figure}[h!]
\includegraphics[width = \textwidth]{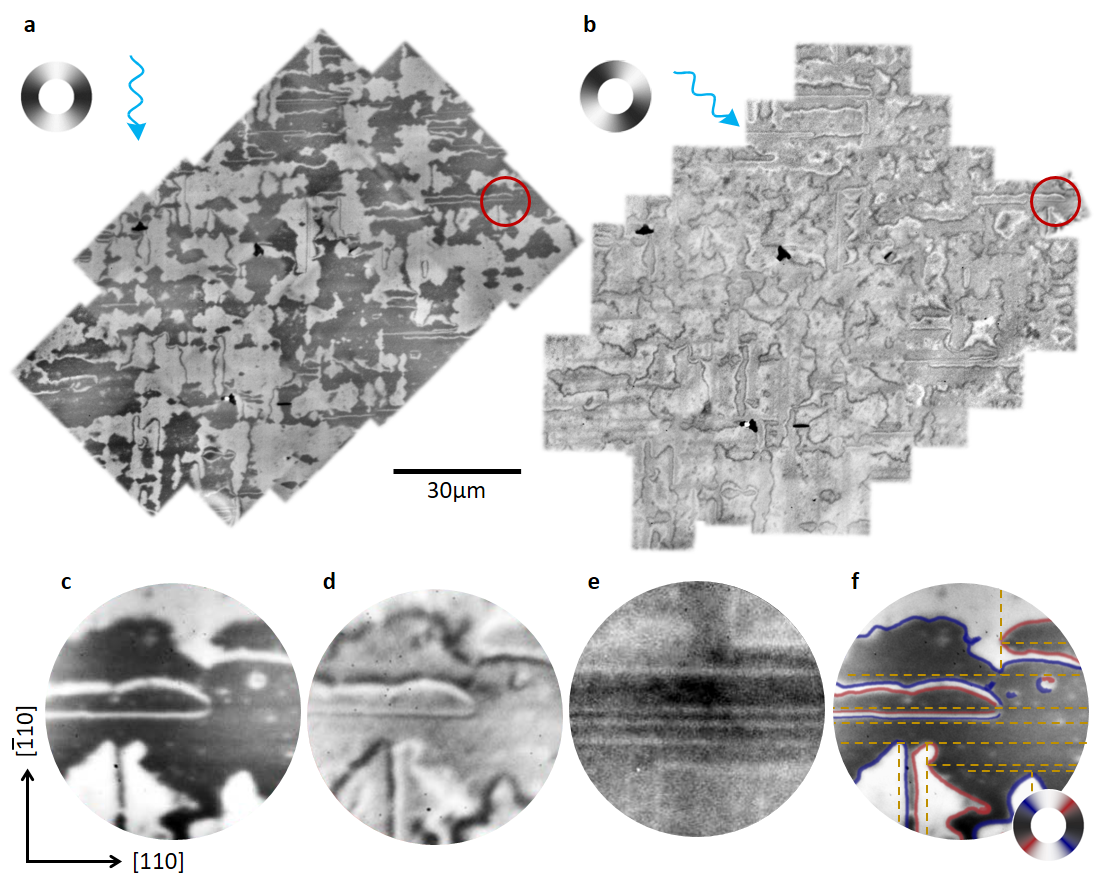}
\caption{ \textbf{AF domains and domain walls.} \textbf{a}, XMLD-PEEM image 
showing AF domains with spin axes parallel to $[110]$ (dark areas) or $[\bar{1}10]$ (light areas). 
\textbf{b}, XMLD-PEEM image for the same area as (a) showing AF domain walls with spin axes parallel to $[100]$ (black lines) or $[010]$ (white lines); the contrast in the domains appears grey. The greyscale wheels indicate the local spin axis in each image and the blue arrows represent the incident X-ray direction. 
The diameter of the red circle is \SI{12.5}{\micro m}.
\textbf{c}, High-resolution XMLD-PEEM image of the domains from the circled area in (a).
\textbf{d}, High-resolution XMLD-PEEM image of the domain walls from the circled area in (b).
\textbf{e}, High-resolution XLD-PEEM image of the same area as in (c) and (d) showing lines arising from defects.
\textbf{f}, Composite image showing the relationship between the XMLD-PEEM and XLD-PEEM images. 
The black and white areas show the magnetic domains, solid red and blue lines show the domain walls with the orientation indicated by the color wheel and the broken yellow lines represent the structural defects revealed by XLD-PEEM.}
\label{fig:OpenSpace}
\end{figure}

The AF domain structure was imaged using high-resolution photoemission electron 
microscopy (PEEM) combined with X-ray magnetic linear dichroism (XMLD) \cite{Wadley17}.
Figures~\ref{fig:OpenSpace}\,\textbf{a} and \ref{fig:OpenSpace}\,\textbf{b} show large area maps imaged with the X-ray polarization vector ($\mathbf{E}$) aligned to highlight the AF domain structure and domain boundaries, respectively. Maximum XMLD contrast is observed between regions with the local spin axis aligned perpendicular and parallel to $\mathbf{E}$. We observe approximately equal populations of light and dark areas 
in Fig.~\ref{fig:OpenSpace}\,\textbf{a}, corresponding to domains with the local spin axis parallel to 
$[110]$ or $[\bar{1}10]$ (see Supplementary Note 1). Figures~\ref{fig:OpenSpace}\,\textbf{c} and \textbf{d} show high-resolution XMLD-PEEM images of the red circled area in Fig.~\ref{fig:OpenSpace}\,\textbf{a} and \textbf{b}. The AF domains typically exceed several \si{\micro \meter} in lateral size and generally have serrated edges. 

The boundaries between the domains are visible in Fig.~\ref{fig:OpenSpace}\,\textbf{b} and \textbf{d}, when $\mathbf{E}$ is at \SI{45}{\degree} to the local spin axis in both AF domains. In this imaging configuration, the domains have the same contrast and the domain boundary contrast dominates. \SI{90}{\degree} domain walls appear as well separated black or white lines, depending on the average direction of the spin axis across the domain wall, with typical width (400$\pm$50)~nm. Adjacent black and white lines in Fig.~\ref{fig:OpenSpace}\,\textbf{b} and \textbf{d} correspond to \SI{180}{\degree} domain walls (see Supplementary Note 2).

X-ray Linear Dichroism (XLD) combined with PEEM is sensitive to local changes in
the charge anisotropy and can therefore act as an indicator of local
crystallographic variations, $i.e.$ structural defects.  Figure~\ref{fig:OpenSpace}\,e shows
an XLD-PEEM image of the red circled area in Fig.~\ref{fig:OpenSpace}\,\textbf{a} and 
\textbf{b} which reveals a pattern of thin lines running parallel to the $[110]$ and $[\bar{1}10]$ crystallographic directions.
Figure~\ref{fig:OpenSpace}\,f shows the AF domain structure superimposed with the 
domain wall contrast (blue and red lines) along with the structural 
defect pattern (broken yellow lines). Direct comparison of the XMLD-PEEM and XLD-PEEM images over the same area shows that the local AF spin axis is always oriented collinear with the defect lines.

Long straight \SI{180}{\degree} domain walls are found to be confined between two parallel defects. These domain walls extend over several microns and can be seen as the long, thin light and dark lines in Fig.~\ref{fig:OpenSpace}\,\textbf{a}. In some cases these domain walls become highly constricted between two neighboring defects as seen in the middle of Fig.~\ref{fig:OpenSpace}\,\textbf{f}.
The \SI{90}{\degree} domain walls form corners in areas where two defects are orthogonal, as for example in the 
bottom half of Fig.~\ref{fig:OpenSpace}\,\textbf{f}, which form the serrated edges seen in Fig.~\ref{fig:OpenSpace}\,\textbf{a}.

\begin{figure}[h!]
\includegraphics [width = 0.8 \textwidth]{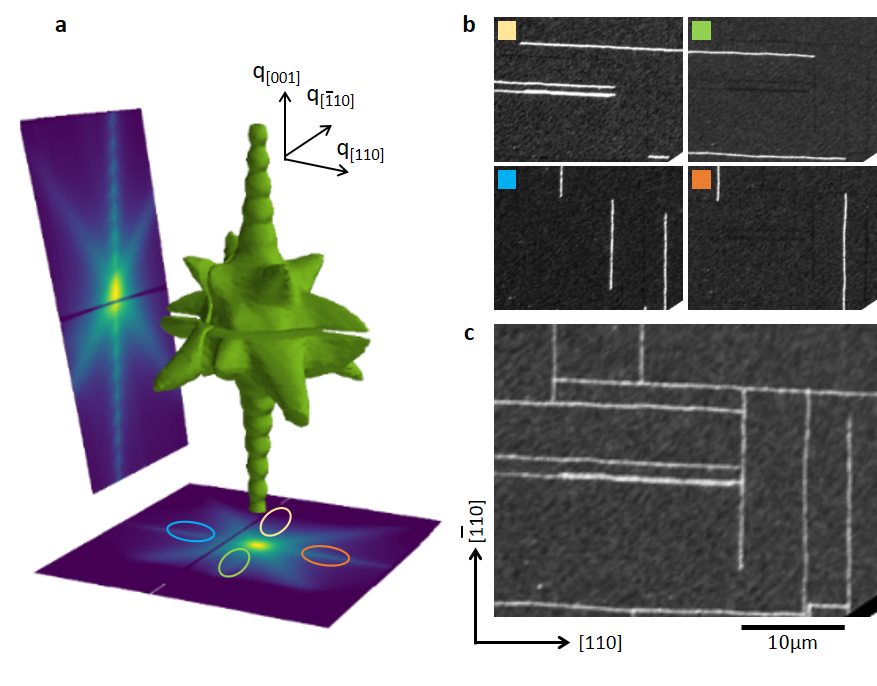}
\caption{\textbf{Bulk microtwinning projected onto the CuMnAs film surface.} \textbf{a}, (003) RSM isosurface (green solid) and projections along the $q_{(110)}$ (left panel) and $q_{(001)}$ (bottom panel) planes. \textbf{b}, Real space SXDM maps extracted from the wings. The color code indicates the areas of reciprocal space used which are given by the corresponding colored ovals in (a). \textbf{c}, Sum of the four images shown in (b). }\label{fig:SXRD} 
\end{figure}

Bulk-sensitive crystallographic information on the nature of defects in thin films,  with nanoscale spatial resolution, can be achieved using scanning X-ray diffraction microscopy (SXDM) \cite{DinaNanobeam}. Crystallographic defects lead to specific contributions to a reciprocal space map (RSM). A three-dimensional RSM of the CuMnAs (003) Bragg peak, generated from two-dimensional diffraction images for several sample tilts (see Methods and Supplementary Note 3) is shown in Fig.~\ref{fig:SXRD}\,\textbf{a}. The RSM has a modulated intensity along $q_{[001]}$ arising from the finite film thickness \cite{pietsch2004high} as well as strong diffuse scattering along the  $q_{\langle 101\rangle}$-type directions, which has been attributed to anti-phase boundaries along the $\{011\}$ planes \cite{Krizek2020}. Sharper intensity streaks, hereafter referred to as wings, along the $q_{\langle 110\rangle}$-type directions indicate the presence of another type of defect. These wings are only visible for specific areas of the sample and are marked by the colored ovals in the lower panel of Fig.~\ref{fig:SXRD}\,\textbf{a}.  

Mapping the intensity of these wings generates real-space images of the defects. 
Maps produced using the intensity of one of the wings yield bright defect lines, on a homogeneous background, which run perpendicular to the direction of the selected wing, see Fig.~\ref{fig:SXRD}\,\textbf{b}. The lines obtained from separate wings are complementary to each other, \textit{i.e.} each wing produces a separate set of lines.
The collated lines arising from all wings are given in Fig.~\ref{fig:SXRD}\,\textbf{c} which reveals a rectangular pattern of defect lines running along the $[110]$ and  $[\bar{1}10]$ directions, reminiscent of the pattern revealed by XLD-PEEM in Fig.~\ref{fig:OpenSpace}\,\textbf{e}. 
The four complementary sets of defect lines with specific $q$-dependence of the scattering revealed by SXDM indicate defect orientations along four different crystallographic directions in the bulk, while the XLD-PEEM images only show their surface termination. 

\begin{figure}[H]
\begin{center}
\includegraphics[width = 0.8 \textwidth]{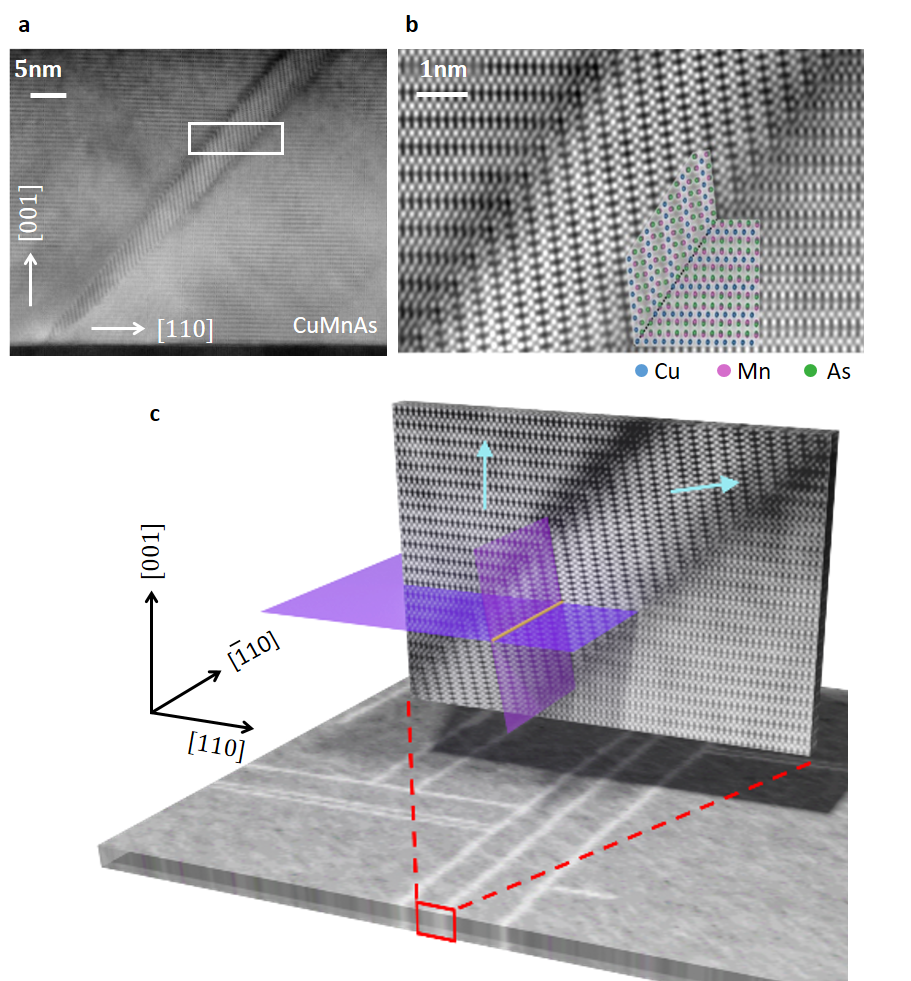}
\caption{\textbf{Atomic structure of the microtwin defects.} \textbf{a}, HAADF-STEM image of a microtwin defect in a CuMnAs thin film. \textbf{b}, HAADF-STEM image from the area shown by the white rectangle in (a), with an atomic model overlay. The microtwin and surrounding bulk film form a coherent boundary indicated by the grey line. \textbf{c}, SXDM map of defect lines on the CuMnAs(001) surface (horizontal panel) and HAADF-STEM image (vertical plane) of a microtwin defect. Teal arrows give the local $c$-axis orientation. Purple sheets indicate the magnetic easy planes in the microtwin and the surrounding film. The orange line 
shows the intersection of the magnetic easy planes which determines the local spin axis for the microtwin and surrounding area.}
\label{fig:Mechanism} 
\end{center}
\end{figure}

High-angle annular dark field-scanning transmission electron microscopy (HAADF-STEM) reveals that the defects are slabs of a microtwinned phase, in which the lattice is rotated so that the $c$-axis is tilted away from the film normal by $\sim \SI{82}{\degree}$, as shown in Fig.~\ref{fig:Mechanism}\,a and b. The slabs extend over most of the film thickness and grow wider towards the sample surface where they produce the characteristic rectangular pattern with lines running parallel to the $[110]$ and $[\bar110]$ directions, Fig.~\ref{fig:Mechanism}\,\textbf{c}. Figure ~\ref{fig:Mechanism}\,\textbf{b} shows a high-resolution image of a microtwin defect where the atomic ordering is indicated. The microtwin and surrounding bulk film form a coherent boundary, with the microtwin slab extending 
along one of the  $\{ 111\}$ planes. In particular, for each defect line on the surface there are two possible bulk defect slab orientations with opposite tilts \cite{Krizek2020} which can be distinguished in SXDM, but not in XLD-PEEM. 

As the magnetic easy-plane in tetragonal CuMnAs is perpendicular to the $c$-axis \cite{wadley15}, the $\sim \SI{82}{\degree}$ rotation of the $c$-axis in the microtwin region will have a profound effect on the local spin orientation. The microtwin region and surrounding bulk film share only one magnetic easy axis, determined by the intersection of the easy planes (purple sheets in Fig.~\ref{fig:Mechanism}\,\textbf{c}). This easy axis is represented by the orange line in Fig.~\ref{fig:Mechanism}\,\textbf{c}.
For any microtwin defect line on the surface, there are two possible propagation directions into the bulk, but for either case the easy axis remains parallel to the defect line on the surface. The local N\'eel vectors (\textit{i.e.} the difference in the sublattice magnetic moment directions) then aligns parallel to the microtwin surface termination. For adjacent microtwin defects, the local N\'eel vector can align either parallel or antiparallel. Antiparallel alignment results in the \SI{180}{\degree} domain walls seen in the XMLD-PEEM images in Fig.~\ref{fig:OpenSpace}. For parallel alignment of the N\'eel vector, the area between the microtwin defects is magnetically homogeneous and can extend over several microns. Perpendicular alignment of two defects gives rise to \SI{90}{\degree} domain walls.

In the final part, we show that including the effects of microtwin defects in micromagnetic simulations is sufficient to fully explain the experimentally observed 
AF domain structures shown in Fig.~\ref{fig:OpenSpace}.
The simulations consider an AF layer with two (equivalent) orthogonal in-plane easy axes, \textit{i.e.} without considering an out-of-plane variation of the N\'eel vector. The effects of microtwins are included as a rotation of the magnetocrystalline anisotropy localized at the microtwins along with a homogeneous strain field, perpendicular to the microtwins (see Supplementary Note 4). We consider that the extrinsic strain due to the microtwin is much larger than the spontaneous strains due to the magnetic texture, such that the inverse effect of the magnetic texture on the strain distribution \cite{Gomonay2002a} can be neglected. The simulations show that the spin axis in the vicinity of a microtwin defect line is always aligned parallel to the defect line.

\begin{figure}[t]
\includegraphics[width = \textwidth]{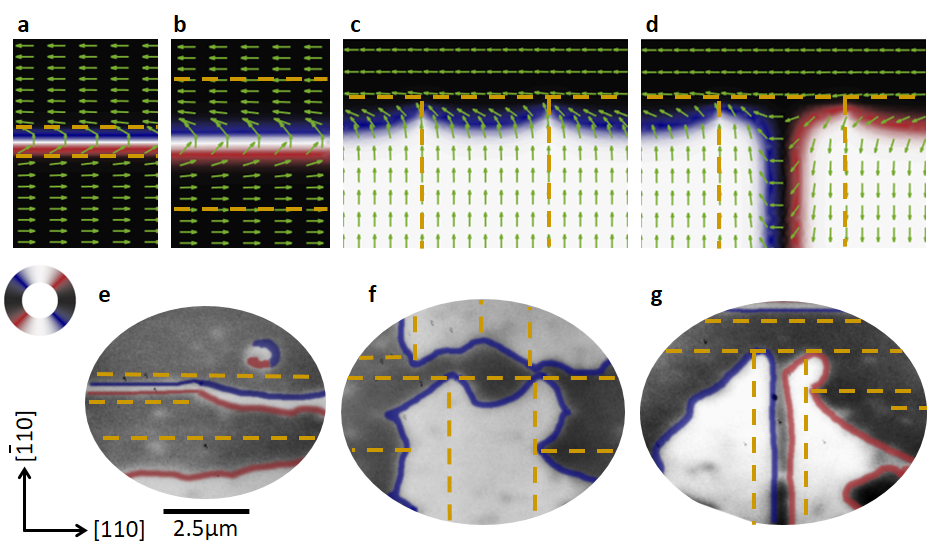}
\caption{\textbf{Micromagnetic simulations.} \textbf{a}-\textbf{d}, Micromagnetic simulations of the AF domain structure in areas with different microtwin patterns (indicated by the broken yellow lines).  \textbf{a}, and  \textbf{b} feature parallel microtwins 2  domain wall widths apart (a) and two 7 domain wall widths apart (b).  \textbf{c}  and \textbf{d} show the simulation results for different initial conditions for the same microtwin pattern of one perpendicular and two parallel microtwins forming two T-junctions. The green arrows (in \textbf{a}-\textbf{d}) and color wheel show the local orientation of the N\'eel vector. \textbf{e}-\textbf{g}, XMLD-PEEM images of AF domains overlaid with the microtwin pattern measured in XLD-PEEM (yellow broken lines). }
\label{fig:Simulations} 
\end{figure}

Figures~\ref{fig:Simulations}\,\textbf{a} and \textbf{b} show the simulated AF domain structures for two parallel microtwin defect lines with a different spacing between the lines.
For this configuration, antiparallel alignment of the N\'eel vectors on either side of the microtwin defects results in a \SI{180}{\degree} domain wall.
For large separations of the microtwins, the domain wall width is determined by the strain-induced magnetic anisotropy. For small  separations, the rotated magnetocrystalline anisotropy of the microtwin defects leads to a narrower, highly confined domain wall. The simulations show a close resemblance to the XMLD-PEEM image of the \SI{180}{\degree} AF domain wall shown in the middle of Fig.~\ref{fig:Simulations}\,\textbf{e}. The domain wall is narrow and straight in the confined area on the left of Fig.~\ref{fig:Simulations}\,\textbf{e} whereas it becomes wider and meanders on the right of the image where the microtwin defects are much further apart.


Figures~\ref{fig:Simulations}\,\textbf{c} and \textbf{d} show simulated AF domain structures for two parallel microtwin defect lines terminating on a perpendicular mircotwin defect line for two different initial boundary configurations. For a parallel boundary configuration of the N\'eel vectors across the two parallel defect lines (Fig.~\ref{fig:Simulations}\,\textbf{c}), the simulations converge to a homogeneous AF domain configuration across the defect lines (see lower half of Fig.~\ref{fig:Simulations}\,\textbf{c}). On the other hand, a \SI{90}{\degree} domain wall forms close to the perpendicular defect line with characteristic pinch points close to the T-junctions reminiscent of the serrated edges seen on the AF domains in the experiment (Fig.~\ref{fig:Simulations}\,\textbf{f}). A similar result is found for an antiparallel initial boundary configuration, but with an additional \SI{180}{\degree} domain wall between the parallel defect lines. This \SI{180}{\degree} domain wall unzips into two \SI{90}{\degree} domain walls with opposite senses of rotation along with pinch points at the T-junctions. The simulation closely reproduces the AF configuration imaged using XMLD-PEEM (Fig.~\ref{fig:Simulations}\,\textbf{g}).

\section{Discussion}
The microscopic AF domain structure in CuMnAs thin films has been shown to be dominated by microtwin defects that appear as lines along specific crystallographic directions on the film surface. The local spin  axis is aligned parallel to the defect lines. This then leads to either large AF domains with serrated edges or \SI{180}{\degree} domain walls running parallel to the $[110]$ or $[\bar{1}10]$ crystallographic directions. A perpendicular orientation of two microtwin defect lines leads to \SI{90}{\degree} domains walls that lead to domain boundaries with serrated edges. The micromagnetic simulations, with the inclusion of local strain-fields arising from the microtwin defects, reproduce all the principal building blocks of the entire AF domain structure observed in CuMnAs thin films. Furthermore, the simulations highlight the variety of domain and domain wall structures possible when an array of microtwin defects govern the AF ordering.

Our results emphasize the sensitivity of  AF domain structures to defect structures in thin films. The concentration of such defects may be engineered by varying the growth parameters, particularly the substrate temperature and Cu/Mn stoichiometry \cite{Krizek2020}, providing a mechanism for tailoring the AF domains and domain walls for specific functionality. Introducing some bias during growth, such as substrate stress or miscut angle, may allow further control over the distribution of defect orientations. In terms of optimization of device performance, it was shown previously that the largest electrical readout signals after current pulsing are observed for growth conditions corresponding to the lowest defect densities \cite{Krizek2020}. This is an indirect indication that the microtwin defects inhibit the nucleation and switching of antiferromagnetic domains. On the other hand, the control of microtwin defects may be beneficial for current-driven motion of antiferromagnetic domain walls between well-defined pinning centres \cite{Wadley2018}. Thus, developing high-performance spintronic devices will rely on a detailed understanding of the nanoscale coupling between the local AF order and the crystallographic microstructure.


\section{Methods}
\subsection{Sample fabrication}
The \SI{50}{\nano \meter} CuMnAs films were grown by molecular beam epitaxy on a GaP buffer layer on a GaP(001) substrate 
and capped with a \SI{3}{\nano \meter} Al film to prevent surface oxidation for the PEEM imaging.

\subsection{PEEM Imaging}
The PEEM measurements were performed on beamline I06 at Diamond Light Source, using linearly polarized X-rays incident at \SI{16}{\degree} to the film surface. Magnetic contrast with $\sim \SI{30}{\nano \meter}$ spatial resolution was obtained from the asymmetry between images recorded using photon energies corresponding to the maximum and  minimum of the Mn $\mathrm{L}_3$ XMLD spectrum (see Supplementary Note 1). The linear polarization ($\mathbf{E}$) was in the plane of the film, and the largest magnetic contrast was obtained between AF domains with spin axes parallel and perpendicular to $\mathbf{E}$. The XMLD spectrum has a a similar lineshape, but opposite sign, for $\mathbf{E} \parallel [110]$ and $\mathbf{E} \parallel [010]$ (Ref. \cite{Hills2015}).
PEEM images with sensitivity to the microtwin configuration were obtained from the asymmetry between images recorded at photon energies corresponding to the maximum and minimum of the Mn $\mathrm{L}_3$ non-magnetic XLD spectrum (see Supplementary Note 1) with $\mathbf{E}$ at \SI{74}{\degree} to the surface. All measurements were performed at room temperature. 

\subsection{Scanning X-ray diffraction microscopy}
SXDM was performed on the \texttt{NanoMAX}-beamline at \texttt{MAX IV Laboratory} \cite{Carbone2020}. The beam was focused to a lateral diameter of \SI{100}{\nano \meter} and the X-ray energy tuned to \SI{10}{keV}.
The measurement geometry was determined by three angles: the detector angle ($\delta$) measured from the direction of the incident beam in the vertical plane, $\Theta$ which defines the angle of incidence with respect to the sample surface in the vertical diffraction plane and $\phi$ which defines the sample azimuth.
The sample was scanned laterally over a 2D-mesh at a fixed sample orientation with a diffraction image recorded at each position. The imaging was performed using a \texttt{Merlin Si Quad} area detector with $512\times 512$ pixels, each $55\times\SI{55}{\micro m}$ in size. The distance between the detector and sample was  \SI{0.65}{\meter}.

RSMs were constructed via several 2D-mesh scans with a stepsize of \SI{200}{\nano \meter} at different $\Theta$ angles around the (003) Bragg reflection in $\SI{0.02}{\degree}$ increments. For these measurements $\phi$ was chosen such that the X-ray beam impinged along the CuMnAs $[110]$ direction. Analysis was performed using the \texttt{xrayutilities} toolbox described in reference \cite{Kriegner2013}.

SXDM imaging of the microtwin configuration was performed  with $\phi$ chosen so that the beam impinged along the CuMnAs $[100]$ direction. 
The microtwin configuration was mapped with the sample at an angle $\Delta \Theta = \pm \SI{0.4}{\degree}$ from the Bragg angle.  
For each angle, the detector plane sliced through two of the microtwin-related wings in reciprocal space. Consequently, if a microtwin was in the illuminated area, significantly higher intensity was recorded on the corresponding area of the detector, depending on the microtwin orientation. 
Mapping these areas of high intensity thus revealed the spatial pattern of microtwins with a specific orientation. The SXDM map shown in Fig.~\ref{fig:SXRD}\,c is  the sum of SXDM images recorded with $\Theta_{\mathrm{Bragg}}-\Delta \Theta$ and $\Theta_{\mathrm{Bragg}} + \Delta \Theta$.
For details see Supplementary Note 3.

\subsection{Transmission Electron Microscopy}
For the HAADF-STEM measurements, the CuMnAs samples were capped with an additional \num{10}-\SI{15}{nm} of carbon \textit{ex situ} and tungsten \textit{in situ}. Thin lamellae were prepared by a Ga focused ion beam. The lamellae were polished at \SI{2}{kV} and \SI{25}{pA}. The lamellae were investigated using a \texttt{FEI Titan Themis 60-300 cubed} high-resolution (scanning) transmission electron microscope at \SI{300}{kV}.
The atomic model overlay shown in Fig.~\ref{fig:Mechanism}\,b was produced using VESTA software \cite{Vesta}.

\subsection{Micromagnetic Simulations}\label{sec:MethodsSim}
The distribution of the N\'eel vector $\mathbf{n}(x,y)$ in the presence of different microtwin configurations was simulated using the \texttt{Matlab PDE Toolbox} to solve  
the micromagnetic equation
\begin{equation}\label{equilibrium_equation}
\mathbf{n}\times\left[A\nabla^2 \mathbf{n}+\mathbf{H}_\mathbf{n}\right]=0
\end{equation}
with von Neumann boundary conditions.
Here $A$ is the magnetic stiffness, $\nabla^2$ is the Laplace operator, and 
\begin{equation}\label{definition_Hn}
\mathbf{H}_\mathbf{n}=-\frac{\partial }{\partial \mathbf{n}}\left(w_\mathrm{an}+w_\mathrm{tw}+w_\mathrm{me}\right),
\end{equation}
where $w_\mathrm{an}$, $w_\mathrm{tw}$ and $w_\mathrm{me}$ are the densities of magnetic anisotropy energy of the bulk film, the magnetic anisotropy energy of the microtwin, and the magnetoelastic energy, respectively. Details can be found in Supplementary Note 4.

\bibliography{references}

\section{Acknowledgements}
We thank Diamond Light Source for the allocation of beamtime on beamline \texttt{I06} under Proposal nos. MM22437-1 and NT27146-1. We acknowledge MAX IV Laboratory for beamtime on Beamline \texttt{NanoMAX} under Proposal C20190533. Research conducted at MAX IV, a Swedish national user facility, is supported by the Swedish Research Council under contract 2018-07152, the Swedish Governmental Agency for Innovation Systems under contract 2018-04969, and Formas under contract 2019-02496. CzechNanoLab project LM2018110 funded by MEYS CR is gratefully acknowledged for the financial support of the measurements and sample fabrication at CEITEC Nano Research Infrastructure. The work was supported by the EU FET Open RIA Grant no 766566 and the Ministry of Education of the Czech Republic Grant No. LM2018110, and LNSM-LNSpin, and the Grant Agency of the Czech Republic Grant No. 19-28375X. 

\section{Author contributions}
SSD, KWE and PW conceived and led the project. SSD, KWE and SR devised the XMLD-PEEM and XLD-PEEM imaging and performed the measurements with FM, OJA, LXB, SFP, KAO and PW. The SXDM experiment were performed by SR, DK, AB, KWE, SSD. DK and DC coordinated and supervised the SXDM data analysis and the interpretation of the results with help from SR. OG developed the micromagnetic simulations with feedback from SSD, KWE and SR. FK, JM and OM performed the HAADF-STEM experiments and data analysis. SR combined and led the analysis of the data from the different experimental techniques. RPC, FK and VN fabricated the samples. SSD, KWE and SR wrote the manuscript with feedback from all authors.

\section{Competing Interests}
The authors declare no competing interests.


\end{document}